\documentclass[aps,pra,showpacs,superscriptaddress,amsfonts,amssymb,amsmath]{revtex4}

\begin{document}

\preprint{Yukawa Institute Kyoto} \preprint{YITP-07-21}

\title{Quasi-exactly solvable Fokker-Planck equations}

\author{Choon-Lin Ho}
 \affiliation{Department of Physics, Tamkang University,
 Tamsui 251, Taiwan, Republic of China}
\author{Ryu Sasaki}
 \affiliation{Yukawa Institute for Theoretical Physics,
     Kyoto University, Kyoto 606-8502, Japan}

\date{Oct 4, 2007} 

\begin{abstract}

We consider exact and quasi-exact solvability of the
one-dimensional Fokker-Planck equation based on the connection
between the Fokker-Planck equation and the Schr\"odinger equation.
A unified consideration of these two types of solvability is given
from the viewpoint of prepotential together with Bethe ansatz
equations.  Quasi-exactly solvable Fokker-Planck equations related
to the $sl(2)$-based systems in Turbiner's classification are
listed.  We also present one $sl(2)$-based example which is not
listed in Turbiner's scheme.

\end{abstract}


 \maketitle

\section{Introduction}

Without doubt the Fokker-Planck (FP) equation is one of the basic tools
used to deal with fluctuations in various kinds of systems
\cite{FP}.  It has found applications in such diverse areas as
physics, astrophysics, chemistry, biology, finance, etc.  Owing to
its wide applicability, various methods of finding exact and
approximate solutions of the FP equations have been developed.

One of the methods of solving the FP equation is to transform the
FP equation into a Schr\"odinger-like equation, and then solve the
eigenvalue problem of the latter.  This method, to be called the method
of eigenfunction expansion, is useful when the associated
Schr\"odinger equation is exactly solvable; for example with infinite
square well, harmonic oscillator potentials, etc.
Several FP equations have been exactly solved in this way \cite{FP}.

Unfortunately exactly solvable Schr\"odinger equations are rather
limited in number, and hence the method of eigenfunction expansion
is quite restricted.  However, in real calculations more often
than not a knowledge of a large but finite number of
eigenfunctions is sufficient for a good approximate result. This
being so, it would render the method of eigenfunction expansion more
useful.  Recently, in non-relativistic quantum mechanics a new
class of potentials which are intermediate to exactly solvable
ones and non-solvable ones has been found. These are called
quasi-exactly solvable (QES) problems for which it is possible to
determine algebraically a part of the spectrum (eigenvalues and
eigenfunctions) but not the whole spectrum \cite{TU,Tur,GKO,Tur3,Ush}.
These QES Schr\"odinger equations thus extend the range of
applicability of the eigenfunction expansion.

In this paper, we shall classify $sl(2)$-based QES FP equations
according to their associated QES Schr\"odinger equations  listed
in \cite{Tur}, and present a case which is not listed in
\cite{Tur}. We first discuss the connection between the FP and
Schr\"odinger equations in Section II.  In Section III the basics
of the exact and quasi-exact solvability of Schr\"odinger
equations are reviewed from the viewpoint of prepotential together
with Bethe ansatz equations, and the way to construct the
corresponding QES FP equations is presented.  Some examples of QES
FP equations constructed in this manner are discussed in Section
IV. Section V concludes the paper.

\medskip

\section{Fokker-Planck and Schr\"odinger equations}

In one dimension, the FP equation of the probability density
$P(x,t)$ is \cite{FP}
\begin{gather}
\frac{\partial}{\partial t} P(x,t)=\mathcal{L}P(x,t),\nonumber\\
\mathcal{L}\equiv -\frac{\partial}{\partial x} D^{(1)}(x) +
\frac{\partial^2}{\partial x^2}D^{(2)}(x). \label{FPE}
\end{gather}
The functions $D^{(1)}(x)$ and $D^{(2)}(x)$ in the FP operator
$\mathcal{L}$ are, respectively,  the drift and the diffusion
coefficient (we consider only time-independent case).  The drift
coefficient represents the external force acting on the particle,
while the diffusion coefficient accounts for the effect of
fluctuation.  The drift coefficient is usually expressed in terms
of a drift potential $\Phi(x)$ according to
$D^{(1)}(x)=-\Phi^\prime (x)$, where the prime denotes the derivative
with respect to $x$.  Without loss of generality, in what follows
we shall take $D^{(2)}=1$.  The stationary solution of the FP
equation is $P_0(x)=\exp(-\Phi(x))$, with $\int P_0(x)\,dx=1$.

The FP equation is closely related to the Schr\"odinger equation
\cite{FP}.  To see this, let us define $\psi(x,t)\equiv
e^{\Phi/2}P(x,t)$.  Substituting this into the FP equation, we
find that $\psi$ satisfies the Schr\"odinger-like equation:
\begin{equation}
\frac{\partial\psi}{\partial t}= \left(\frac{\partial^2}{\partial
x^2} -
\left(\frac{\Phi^\prime}{2}\right)^2+\left(\frac{\Phi}{2}\right)
^{\prime\prime}\right)\psi. \label{S-like}
\end{equation}
As we shall see in what follows, the function $\Phi/2$ plays a
fundamental role in our discussions.  So let it be denoted by
$W(x)\equiv \Phi(x)/2$.  Following \cite{ST}, we shall call $W(x)$
the prepotential.  Letting $\psi(x,t)=\exp(-\lambda t)\phi(x)$ and
$\phi_0\equiv \exp(-W)$, we obtain
\begin{equation}
H\phi=\lambda\phi,
\end{equation}
where
\begin{eqnarray}
H&\equiv& -\phi_0^{-1}\mathcal{L}\phi_0 \nonumber\\
&=&-\frac{\partial^2}{\partial x^2}+V(x),\\
V(x)&=&W^\prime (x)^2 - W^{\prime\prime}(x).\nonumber
\end{eqnarray}
Thus $\phi$ satisfies the time-independent Schr\"odinger equation with
Hamiltonian $H$ and eigenvalue $\lambda$, and $\phi_0$ is the zero
mode of $H$: $H\phi_0=0$. Since only derivatives of $W(x)$ appear
in $V(x)$, $W(x)$ is defined only up to an additive constant.  We
choose the constant in such a way as to normalize $\phi_0(x)$
properly, $\int\phi_0(x)^2\,dx=1$. For simplicity of presentation,
we consider the cases in which the ground state wave functions are
square integrable, that is the corresponding FP operators have the
normalizable stationary distribution.

Now comes the essence of the method of eigenfunction expansion for
solving FP equations. If all the eigenfunctions $\phi_n$
($n=0,1,2,\ldots$) of $H$ with eigenvalues $\lambda_n$ are solved,
then the eigenfunctions $P_n(x)$ of $\mathcal{L}$ corresponding to
the eigenvalue $-\lambda_n$ is $P_n(x)=\phi_0(x)\phi_n(x)$.  The
stationary distribution is $P_0=\phi_0^2=\exp(-2W)$, which is
obviously non-negative, and is the zero mode of $\mathcal{L}$:
$\mathcal{L}P_0=0$. Any positive definite initial probability
density $P(x,0)$ can be expanded as $P(x,0)=\phi_0(x)\sum_n
c_n\phi_n(x)$, with constant coefficients $c_n$ ($n=0,1,\ldots$)
\begin{eqnarray}
c_n=\int_{-\infty}^\infty \phi_n(x)\left(\phi_0^{-1}(x)
P(x,0)\right)dx.
\end{eqnarray}
Then at any later time $t$, the solution of the FP equation is
$P(x,t)=\phi_0(x)\sum_n c_n \phi_n(x)\exp(-\lambda_n t)$.

It is now evident that the exact and quasi-exact solvability of FP
equations are directly related to the exact and quasi-exact
solvability of the corresponding Schr\"odinger equations.

We shall classify the types of exactly solvable and QES  FP
equations associated with Schr\"odinger equations which are
factorizable (or supersymmetric).   For this, it is desirable to
have a unified approach to deal with both kinds of solvability. It
appears to us that the approach based on the so-called
prepotential serves this purpose well.  We shall therefore first
review exact and quasi-exact solvability from this viewpoint
below.

\section{Exact and quasi-exact solvability from the point of view
of prepotential with Bethe ansatz equations}

Let us now briefly review the essence of quasi-exact and exact
solvability of Schr\"odinger equations.  We choose to proceed from
the point of view of prepotentials together with the Bethe ansatz
equations \cite{ST} (see also \cite{Ho}).  The merit of this
approach is that one needs not to first assume the form of the
potential of the system concerned, as all information about the
system is contained in the prepotential and the solutions, or
roots, of the Bethe ansatz equations \cite{Bethe}.   Also, in this
approach exact and quasi-exact solvability can be treated on the
same footing.  Furthermore, such approach facilitates extension of
the QES theory from the Schr\"odinger equation to equations for
multi-component wave functions, such as the Pauli and the Dirac
equation \cite{Ho}.

Suppose $\phi_0\equiv e^{-W_0(x)}$ is the ground state, with zero
energy, of a Hamiltonian $H_0$: $H_0\phi_0=0$. This is the simplest
example of quasi-exact solvability. This implies that the
potential $V_0$ is completely determined by $W_0$:
$V_0={W_0^\prime}^2 - W_0^{\prime\prime}$, and consequently,
the Hamiltonian is factorizable:
\begin{equation}
H_0=\left(-\frac{d}{dx}+W_0^\prime\right)\left(\frac{d}{dx}+W_0^\prime\right).
     \label{facHam}
\end{equation}
This fact can also be considered as the very base of
supersymmetric quantum mechanics.  Consider now a wave function
$\phi_N$ ($N$: positive integer) which is related to $\phi_0$ of
$H_0$ by $\phi_N=\phi_0{\tilde \phi}_N$:
\begin{eqnarray}
 {\tilde\phi}_N=(z-z_1)(z-z_2)\cdots
(z-z_N). \label{phi-2}
\end{eqnarray}
 Here $z=z(x)$ is some function of $x$. The function
${\tilde\phi}_N$ is a polynomial in an $(N+1)$-dimensional Hilbert
space with the basis $\langle 1,z,z^2,\ldots,z^N \rangle$.  One
can rewrite $\phi_N$ as
\begin{eqnarray}
 \phi_N =\exp\left(- W_N(x,\{z_k\})
\right), \label{f2} \end{eqnarray} with
\begin{eqnarray}
W_N(x,\{z_k\}) = W_0(x) - \sum_{k=1}^N \ln |z(x)-z_k|. \label{W}
\end{eqnarray}
Now we form  $W_N^{\prime 2} - W_N^{\prime\prime}$.  This results
in a Riccati equation
\begin{eqnarray}
 W_N^{\prime 2} (x,\{z_k\})- W_N^{\prime\prime}(x,\{z_k\})= V_N
 (x,\{z_k\})
- \lambda_N(\{z_k\}), \label{Riccati}
\end{eqnarray}
where $V_N$ is in general a function of $N$, $x$ and $\{z_k\}$,
and $\lambda_N$ is a real constant depending on $N$ and $\{z_k\}$
(dependence of $V_N$ and $\lambda_N$ on other parameters of $H_0$
is understood).  It should be noted that the definition of $V_N$
and $\lambda_N$ has ambiguity, namely, they can both be shifted by
an arbitrary real constant $\alpha$:
\begin{equation}
V_N\to V_N+\alpha,\quad \lambda_N\to \lambda_N+\alpha.
\end{equation}
This corresponds to the redefinition of the zero point of $V_N$.
For the theory we are considering, the r.h.s. of (\ref{Riccati})
is generally a meromorphic function of $z$ with at most simple
poles. Let us demand that the residues of the simple poles,
$z_{k}$, $k=1,\ldots, N$ should all vanish. This will result in a
set of algebraic equations which the parameters $\{z_k\}$ must
satisfy. These equations are called the Bethe ansatz equations for
$\{z_k\}$.  For each value of $N$ ($N=0,1,2,\ldots$), there are
$N+1$ possible sets of solution $\{z_k\}$ of the Bethe ansatz
equations

With $\{z_k\}$ satisfying the Bethe ansatz equations, the r.h.s.
of (\ref{Riccati}) will have no simple poles at $\{z_k\}$ but it
still  generally depends  on $\{z_k\}$.  Suppose now that the
$N+1$ sets of roots give the same form of the potential $V_N
(x,\{z_k\})=V_N (x)$ in the r.h.s. of (\ref{Riccati}), but with
$N+1$ values of $\lambda_N$. Then the Riccati equation
(\ref{Riccati}) implies that we have a new eigenvalue problem with
Hamiltonian $\mathcal{H}_N$:
\begin{gather}
\mathcal{H}_N\equiv -\frac{d^2}{dx^2} + V_N,\nonumber\\
\mathcal{H}_N\phi_N = \lambda_N\phi_N. \label{H_N}
\end{gather}
The $N+1$ sets of $\{z_k\}$ define $N+1$ eigenfunctions $\phi_N$
of $\mathcal{H}_N$ with eigenvalue $\lambda_N$.  In other words,
$N+1$ eigenfunctions and eigenvalues of the Hamiltonian
$\mathcal{H}_N$ are solvable algebraically, and $\mathcal{H}_N$ is
said to be QES.  If the wave function $\phi_N^{(0)}\equiv \exp(-
W_N^{(0)})$ corresponding to the lowest value
$\lambda_N=\lambda_N^{(0)}$ has no node, then it is the ground
state of $\mathcal{H}_N$, and is annihilated by the Hamiltonian
$H_N$:
\begin{gather}
H_N \equiv \mathcal{H}_N-\lambda_N^{(0)}
=-\frac{d^2}{dx^2} + W_N^{(0)\prime 2} - W_N^{(0)\prime\prime},\\
H_N\phi_N^{(0)}=0.
\end{gather}
Now if $V_N\equiv V$ is not only independent of the solutions of
the Bethe ansatz equations $\{z_k\}$, but is also independent of
$N$, then the Hamiltonian $H$ is exactly solvable, as solvable
states can be found for any integral value of $N$.

{}From the Schr\"odinger equation
$\mathcal{H}_N\phi_N=\lambda_N\phi_N$, it is also seen that the
polynomial part ${\tilde\phi}_N$ satisfies the equation
$h_N{\tilde\phi}_N=\lambda_N{\tilde\phi}_N$, where
$h_N\equiv\phi_0^{-1}\mathcal{H}_N\phi_0$, and $\phi_0$ is the
ground state of $H_0$ corresponding to $N=0$. From the previous
discussions, we know that $h_N$ has an algebraic sector with $N+1$
eigenvalues and eigenfunctions, which are polynomials in an
$(N+1)$-dimensional Hilbert space with the basis $\langle
1,z,z^2,\ldots,z^N \rangle$. The quasi-exact solvability of $H_N$
is said to have a Lie-algebraic origin, if $h_N$ can be expressed
as
\begin{eqnarray}
h_N=\sum C_{ab}J^a J^b + \sum C_a J^a + {\rm constant},
\label{H-g}
\end{eqnarray}
where $C_{ab},~C_a$ are constant coefficients, and the $J^a$ are
the generators of some Lie algebra.  QES Hamiltonians based on the
Lie-algebra $sl(2)$ have been classified in \cite{Tur,GKO}.

There exists a different situation when the $N+1$ sets of roots
$\{z_k\}$ are plugged into the l.h.s. of (\ref{Riccati}). Instead
of resulting in the r.h.s. of (\ref{Riccati}) a single potential
$V_N$ with $N+1$ values of $\lambda_N$, $N+1$ potentials $V_N$
with a single value of $\lambda_N$ are obtained, when the zero
point of the eigenvalue is properly adjusted.  These are QES
systems for which a set of $N+1$ potentials differing by the
values of parameters have the same eigenvalue of the $j$-th
eigenstate in the $j$-th potential ($j=1,2,\ldots,N+1$). Such kind
of QES systems were termed the second type QES systems, and those
discussed previously were called the first type in \cite{Tur}. For
the second type QES quantum systems which have $sl(2)$
Lie-algebraic origin, it is $(h_N-E)/\rho (x)$ instead of $h_N$,
where $\rho(x)$ is some $x$-dependent scale factor, that is
expressible in terms of certain quadratic combination of the
generators $J^a$'s. In this sense,  one can say that for the first
type systems, the scale factors $\rho$ are simply some real
constants.

Combining the discussions in this section and those in the preceding
section, we arrive at a recipe of constructing QES FP equation
from a known QES Schr\"odinger equation. We shall be interested
only in the first type QES systems,  since for the second type QES
systems the corresponding $W_N$ only define $N+1$ FP systems with
$N+1$ different drift potentials $\Phi_N=2W_N$ having the same
eigenvalue.  While these are QES FP systems, it is not useful for
good approximate calculations using the method of eigenfunction
expansion as only one eigenstate is known.

Suppose a prepotential $W_0$ and a corresponding set of Bethe
ansatz equations define a first type QES Schr\"odinger equation in
the sense discussed above. Then the drift potential $\Phi_0=2W_0$
defines a QES FP equation with only one exactly solved state
$\phi_0$, and the stationary solution of the FP equation is
$P_0(x)=\phi_0^2$.  To obtain QES FP equations admitting a larger
number of solvable states, we construct a new prepotential $W_N$
from $W_0$ and the $N$ roots $\{z_1,z_2,\ldots,z_N\}$ of the Bethe
ansatz equations according to (\ref{W}).  The set of roots
$\{z_k\}$ to be chosen is the set for which the value of
eigenvalue $\lambda_N$ is the lowest. If the corresponding wave
function $\phi_N^{(0)}$ has no node, one can use it to define a
corresponding FP operator:
\begin{eqnarray}
\mathcal{L}_N\equiv -\phi_N^{(0)} H_N \phi_N^{(0)-1}.
\end{eqnarray}
Then $\mathcal{L}_N$ defines a QES FP equation having $N+1$
solvable states, with a drift potential $\Phi_N=2W_N^{(0)}$.  As
mentioned before, if $H_N$ is independent of $N$, then $H_N$ is
exactly solvable, and so is the corresponding FP equation.

We shall illustrate this procedure by some examples below.

\section{Examples}

\subsection{Exactly solvable cases}

It is clear that FP equations transformable to exactly solvable
Schr\"odinger equations can be exactly solved.  Particularly, all
the shape-invariant potentials in supersymmetric quantum
mechanics, as listed in Table~4.1 of \cite{Cooper}, or
equivalently, Table~5.1 in \cite{Junker}, give the corresponding
exactly solvable FP systems.  One needs only to link the
prepotential $W$ \cite{note} in the Schr\"odinger system with the
drift potential $\Phi$, or drift coefficient $D^{(1)}$, in the FP
system according to the relation $\Phi=2W$ and
$D^{(1)}=-\Phi^\prime$. We shall not list all of them here, but
simply mention that the FP equation for the Ornstein-Uhlenbeck
process corresponds to the shifted oscillator potential in quantum
mechanics, and that for the Rayleigh process corresponds to the
three-dimensional oscillator potential.

\subsection{$sl(2)$-based QES cases listed in Turbiner's classification}

Similarly, one can identify all QES FP equations which are based
on the $sl(2)$ Lie algebra from the list of ten classes depicted
in Table~1 of \cite{Tur}. Of the ten classes,  classes II, III, V,
VIII and IX belong to the second type QES systems which, as
discussed previously, are not useful as far as the method of
eigenfunction expansion is concerned.    Class X gives a periodic
potential, which is also not of concern in this paper, as we are
interested in normalizable wave functions so that the distribution
density of the FP equation is also normalizable.

So we are left with only classes I, IV, VI and VII in the list of
\cite{Tur} which give QES FP systems having $N+1$ eigenstates with
arbitrary $N$.  To obtain these four QES FP systems, one needs
only to identify $W_0$ with the gauge function $g$ of each class
listed in \cite{Tur}.  This gives a QES FP system with only one
solvable eigenstate.  For QES FP systems in the same class but
with higher values of $N$, we construct $W_N$ from $W_0$ and the
roots of the Bethe ansatz equations $\{z_k\}$ (with the function
$z=z(x)$ also from \cite{Tur}) according to the procedure in
\cite{Ho}.

Let us illustrate this by class VII QES FP system. It turns out
that this case is just the one-particle case of the so-called
rational $BC$ type Inozemtsev model \cite{ST}.

In this case $z(x)=x^2$, and the prepotential (\ref{W}) is given
by
\begin{gather}
     W_N (x) = W_{0}+{\bar W},
     \\
     W_{0}={a\over4}x^4
     +{b\over2}x^2-\gamma\ln x + C_N,~x>0, ~ a>0, ~ \gamma>0,
     \label{eq:sexpot}  \\
  {\bar W}=-\sum_{k=1}^{N}\ln|x^2-z_{k}|.
\end{gather}
Here $C_N$ is a constant term necessary for the normalization of
the ground state wave function $\phi^{(0)}_N(x)$.  By plugging in
all $N+1$ possible  solutions $\{z_k\}$ into $V_N$ in
(\ref{Riccati}), we find that $V_N$ is independent of $\{z_k\}$.
The Bethe ansatz equations removing the simple poles at $\{z_k\}$
read
\begin{eqnarray}
2az_k^2 +2bz_k -\left(2\gamma+1\right) - 4\sum_{l\neq
k}\frac{z_k}{z_k-z_l} =0, ~~ k=1,\ldots,N, \label{BA-VII}
\end{eqnarray}
and $\lambda_N$ in terms of the roots $z_k$'s is
\begin{eqnarray}
\lambda_N=2\left(2\gamma+1\right)\sum_{k=1}^N \frac{1}{z_k},
\label{E-VII}
\end{eqnarray}
or equivalently, using the Bethe ansatz equation,
\begin{equation}
\lambda_N= 4bN+4a\sum_{k=1}^N z_k.
\end{equation}
 Among all the
solutions, we choose those $\{z_k\}$ which are all negative
\begin{equation}
z_k<0,\quad k=1,\ldots, N. \label{neg-root}
\end{equation}
This corresponds to the ground state, since the eigenvalue
$\lambda_N^{(0)}$ is the lowest, and the wave function
\begin{equation}
\phi_N^{(0)}=e^{-({a\over4}x^4
     +{b\over2}x^2)}x^\gamma\prod_{k=1}^N(x^2-z_k)
     \label{phi_N}
\end{equation}
has no node.  The fact that the choice (\ref{neg-root}) is
possible can be understood as follows.  Recall that the number of
QES states is $N+1$ for each $N$.  Since the factor
$\prod_{k=1}^N(x^2-z_k)$ in the wave function (\ref{phi_N}) is
even in $x$, there can be at most $N$ zeros in the half line
$x>0$. According to a well known theorem in quantum mechanics, no
two eigenstates can have the same number of nodes.  This implies
that the number of nodes of the $N+1$ QES states here must be
$0,1,2,\ldots,N$.  Hence the ground state is in the QES sector,
and the choice (\ref{neg-root}) is guaranteed. A similar argument
can be applied to the other cases.

The corresponding QES FP equation is then defined by the drift
potential
\begin{equation}
\Phi_N={a\over2}x^4
     +bx^2-2\gamma\ln x-2\sum_{k=1}^{N}\ln |x^2-z_{k}|+{\rm constant}.
\end{equation}
The constant is chosen so that the stationary distribution of this
FP equation  $\phi_N^{(0)2}=\exp(-\Phi_N)$ is normalized. We have
thus obtained a QES FP equation corresponding to the class VII
$sl(2)$-based QES Schr\"odinger equation in \cite{Tur}.

\begin{table}[!]
\caption{\label{table1} Three other QES FP systems associated with
the $sl(2)$-based QES Schr\"odinger equations in \cite{Tur}. The
function $z(x)$, prepotential $W_0 (x)$ ($-\infty<x<\infty$),
Bethe ansatz equations, and the eigenvalues $\lambda_N$ are
listed. The drift potential defining the QES FP equation with
$N+1$ solvable states is $\Phi_N=2W_N$, where the prepotential
$W_N$ is constructed from (\ref{W}) with $W_0$ and the roots of
the Bethe ansatz equations. All parameters are real.}
\begin{ruledtabular}
\begin{tabular}{ccccc}
Class & $z(x)$ & $W_0$ & Bethe ansatz equations ($k=1,2\ldots,N$) & $\lambda_N$\\
 \hline\\
 I & $e^{-\alpha x}$& $\frac{a}{\alpha}e^{-\alpha x} +bx + \frac{c}{\alpha}e^{\alpha
 x}$
 & $az_k^2 -(b+\frac{\alpha}{2})z_k - c $ & $2\alpha c\sum_{k=1}^N
\frac{1}{z_k}$\\
&& ($\alpha>0,~a, c\geq 0$, $\forall ~b$) & $-\alpha \sum_{l\neq
k}\frac{z_k^2}{z_k-z_l} =0$ &
\\ \\
\hline\\
 IV & $\cosh^{-2}\alpha x$ &
$\frac{c}{4\alpha}\cosh2\alpha x + \frac{a}{\alpha}\ln\cosh\alpha
x$ & $(a+\frac{3}{2}\alpha) z_k^2 +(c-a-\alpha)z_k - c $
   & $ 2ac-a^2-\alpha c$ \\
&& ($\alpha>0,~c\geq0,~\forall~a$)&$- 2\alpha \sum_{l\neq
k}\frac{z_k^2(1-z_k)}{z_k-z_l} =0$ &$+4\alpha c\sum_{k=1}^N
\frac{1}{z_k}$
 \\ \\
\hline\\
  VI  & $x^2$ & $\frac{a}{4}x^4 + \frac{b}{2}x^2$ &  $az_k^2 +bz_k - \frac{1}{2}$
   & $2\sum_{k=1}^N \frac{1}{z_k}$\\
&& ($a>0,~\forall~b$ or $a\geq 0,~b>0$) & $-2\sum_{l\neq
k}\frac{z_k}{z_k-z_l} =0$ &
\\ \\
\end{tabular}
\end{ruledtabular}
\end{table}

QES FP equations of the other three classes can be treated
accordingly.  The main defining information is given in Table~1.

\subsection{$sl(2)$-based QES case not listed in Turbiner's classification}

We now present a QES case which is also based on the $sl(2)$
algebra, but not listed in \cite{Tur}.  This is related to the
trigonometric $BC$ type Inozemtsev model with one degree of
freedom \cite{ST}.

The prepotential in this case reads
\begin{align}
W_N&={a\over2}\cos2x
     -{b\over2}\ln |\cot x|-c\log|\sin x| \nonumber\\
     & -\sum_{k=1}^N\ln |\sin^2x-z_{k}|+C_N,
      ~~ 0<x<\frac{\pi}{2}.
      \label{W2}
\end{align}
All the parameters $a,b$ and $c$ are real and satisfy
$a>0,~c>\frac{b}{2}>0$. Again, $C_N$ is a constant term necessary
for the normalization of the ground state wave function
$\phi^{(0)}_N(x)$. The first three terms in (\ref{W2}) are just
$W_0$ which gives only the ground state solution. The parameters
$\{z_k\}$ are determined by the following Bethe ansatz equations:
\begin{gather}
(4az_k+2c)(1-z_k)-b+1-2z_k
+4z_k(1-z_k)\sum_{l\neq k}{1\over{z_k-z_l}}=0,\nonumber\\
 k=1,\ldots,N. \label{sinbet}
\end{gather}
With these solutions, the potential $V_N$ and eigenvalue
$\lambda_N$ in the Riccati equation (\ref{Riccati}) are
\begin{equation}
V_N(x,\{z_k\})=\left(\frac{d
W_{0}}{dx}\right)^2-\frac{d^2W_{0}}{dx^2}-8aN\sin^2x,\label{V2}
\end{equation}
and
\begin{equation}
\lambda_N(\{z_k\})= 8a\sum_{k=1}^N z_{k}+4N(N+c).
     \label{eq:enei2}
\end{equation}
Again  we choose those $\{z_k\}$ which are all negative
\[
z_k<0,\quad k=1,\ldots, N.
\]
This corresponds to the ground state and its wave function has no
node:
\begin{align}
e^{-W_N^{(0)}(x)}\propto e^{-{a\over2}\cos2x}(\cot
x)^{\frac{b}{2}}(\sin x)^c\prod_{k=1}^N(\sin^2x-z_k).
\end{align}
The corresponding QES FP equation is then defined by the drift
potential $\Phi_N=2W_N^{(0)}$.

The prepotential $W_N$ in (\ref{W2}), and hence the potential
$V_N$ in (\ref{V2}) are not listed in the $sl(2)$-based
classification scheme in \cite{Tur}. Nevertheless, it turns out
that this system is also related to $sl(2)$ Lie algebra
\cite{Zhang}. This can be seen as follows.  Using
$\phi_0=\exp(-W_0)$ and $V_N$, we obtain the transformed
Hamiltonian $h_N=\phi_0^{-1}\mathcal{H}_N\phi_0$ as
\begin{eqnarray}
    h_N & = &  -\frac{d^2}{dx^2}-2\frac{d W_{0}}{dx}\frac{d}{dx}-8aN\sin^2x,
\nonumber  \\
      & = & -\frac{d^2}{dx^2}-2\left(a\sin2x-\frac{b}{\sin2x}+c\cot x\right)
\frac{d}{dx} -8aN\sin^2x.
\end{eqnarray}
In terms of the new variable $z(x)=\sin^2 x$, $h_N$ becomes
\begin{equation}
h_N=-4z(1-z)\frac{d^2}{dz^2}+2\left[4az^2 + 2\left(c-2a+1\right)z
+ b-2c-1\right]\frac{d}{dz} - 8aNz.\label{h-z}
\end{equation}
Now it is easy to check that eq.~(\ref{h-z}) can in fact be
rewritten in the form (\ref{H-g}) as
\begin{equation}
h_N=4J^+J^- -4J^0J^- +8aJ^+ +4\left(c-2a+1+N\right)J^0
+2\left(b-2c-1-N\right)J^- + 2N\left(c-2a+1+N\right),
\end{equation}
where the generators of the Lie algebra $sl(2)$ are given by
\begin{eqnarray}
J^+ &=& z^2 \frac{d}{dz} - Nz~,\cr
J^0&=&z\frac{d}{dz}-\frac{N}{2}~,~~~~~~~~N=0,1,2\ldots\cr J^-&=&
\frac{d}{dz}~.
\end{eqnarray}
This proves that the trigonometric $BC$ type Inozemtsev model is
indeed a QES system based on $sl(2)$.

\section{Summary}

In summary, we have discussed exact and quasi-exact solvability of
the FP equation based on the corresponding solvability of its
associated Schr\"odinger equation.  We give a unified treatment of
these two types of solvability from the viewpoint of prepotential
together with Bethe ansatz equations. Examples of QES FP equations
related to the $sl(2)$-based systems listed in \cite{Tur} are
classified, and one example which is also based on the $sl(2)$
algebra but not listed in \cite{Tur} is also presented. We note
here that, as shown in \cite{ST}, the concepts of quasi-exact
solvability and the so-called higher derivative, or nonlinear, or
${\mathcal N}$-{\em fold} supersymmetry \cite{N-SUSY} are
equivalent. Hence, all the QES FP equations discussed here also
possess such kind of extended supersymmetry.

As mentioned before, we only discuss those QES FP systems which
can admit a large number of QES states so that they are useful for
consideration when approximate calculations are required.  Hence,
we have not considered the so-called second type QES systems
defined in \cite{Tur}, as they admit only a single solvable state
and are therefore of less interest in practical calculations.
Nevertheless, these QES quantum systems still give the
corresponding QES FP equations.  We hope that in the future more
non-$sl(2)$ QES FP systems can be found.

\begin{acknowledgments}

This work is supported in part by the National Science Council
(NSC) of the Republic of China under Grant NSC 95-2112-M-032-012
(CLH), and in part by the Grant-in-Aid for Scientific Research
from the Ministry of Education, Culture, Sports, Science and
Technology under Grant No.18340061 (RS). We would like to thank
Koryu-kyokai (Japan) and  National Center for Theoretical Sciences
(Taipei) for support through the Japan-Taiwan collaboration
programs.  CLH and RS would like to thank the staff and members of
the Yukawa Institute for Theoretical Physics, Kyoto University and
the National Taiwan University, respectively, for their
hospitality and financial support during their respective visits.
We thank Yao-Zhong Zhang for pointing out that the trigonometric
$BC$ type Inozemtsev model is related to $sl(2)$ algebra.

\end{acknowledgments}


\begin{thebibliography}{99}
\bibitem{note}
We note here that our prepotential $W$ is the integral of the
superpotetial $W$ in \cite{Cooper,Junker}.
\end{thebibliography}
\end{document}